# Clustered vacancies in ZnO: Chemical aspects and consequences on physical properties

S. Pal[a], N. Gogurla[b], Avishek Das[c], S. S. Singha[d], P. Kumar[e], D. Kanjilal[e], A. Singha[d], S. Chattopadhyay[c], D. Jana[a] and A. Sarkar[f,*]

[a]Department of Physics, University of Calcutta, 92 Acharya Prafulla Chandra Road, Kolkata 700009, India

[b]Department of Physics, Indian Institute of Technology, Kharagpur 721302, India

[c]Department of Electronic science, University of Calcutta, 92 Acharya Prafulla Chandra Road, Kolkata 700009, India

[d]Department of Physics, Bose Institute, 93/1, Acharya Prafulla Chandra Road, Kolkata 700009, India

[e]Inter University Accelerator Centre, Post Box 10502, Aruna Asaf Ali Marg, New Delhi 110067, India

[f]Department of Physics, Bangabasi Morning College, 19 Rajkumar Chakraborty Sarani, Kolkata 700009, India

[*]Corresponding author: Tel. ++ 91 33 2360 7586, e-mail- sarkarcal@gmail.com

**Abstract**

Chemical nature of point defects, their segregation, cluster or complex formation in ZnO is an important area of investigation. In this report, 1.2 MeV Ar ion beam is used to incorporate defects in granular ZnO. Evolution of defective state with irradiation fluence $1 \times 10^{14}$ and $1 \times 10^{16}$ ions/cm$^2$ has been monitored using x-ray photoelectron spectroscopy (XPS), photoluminescence (PL) and Raman spectroscopic study. Choice of such fluence regime ensures spatial overlapping of successive ion-target collision cascades so that reorganization of point defects can take place in the form of more stable clusters or in the direction of possible recovery. XPS study shows presence of oxygen vacancies ($V_O$) in the Ar ion irradiated ZnO. Zn(LMM) Auger spectra clearly identifies transition involving metallic zinc in the irradiated samples. Intense PL emission from interstitial Zn ($I_{Zn}$) related shallow donor bound excitons (DBX) is visible in the 10 K spectra for all samples. Although overall PL is largely reduced with irradiation disorder, DBX intensity is increased for the highest fluence irradiated sample. Raman study indicates damage in both zinc and oxygen sub-lattice by energetic ion beam. Representative Raman modes from defect complexes involving $V_O$, $I_{Zn}$ and $I_O$ are visible after irradiation with intermediate fluence. Further increase of fluence shows, to some extent, a homogenization of disorder. Huge reduction of resistance is also noted for this sample. Certainly, high irradiation fluence induces a qualitative modification of the conventional (and highly resistive) grain boundary (GB) structure of granular ZnO. Low resistive path, involving $I_{Zn}$ related shallow donors, across the GB can be presumed to explain resistance reduction. Open volumes ($V_{Zn}$ and $V_O$) agglomerate more and more with increasing irradiation fluence and finally get transformed to voids. Results as a whole have been elucidated with a model which emphasizes possible evolution of new defect microstructure that is distinctively different from the GB related disorder. Based on the model, qualitative explanations of commonly observed radiation hardness, colouration and ferromagnetism in disordered ZnO have been put forward. A coherent scenario on disorder accumulation in ZnO has been presented, which we believe, will guide further discussion on this topic.

## 1. Introduction

The presence of defects in any material is ubiquitous and ZnO is not an exception in true sense. Extensive theoretical and experimental efforts have been carried out to understand individual point defects in ZnO with specific chemical identity and their respective key role on physical and chemical properties[1-5]. At the same time, efforts to acquire knowledge on extended defects in ZnO such as grain boundary, dislocations also are being continued[6-10]. Sometimes, point defects in the bulk migrate to the defect sink of extended defects present in the material[11]. On another side, significant numbers of studies have revealed the presence of defect clusters or pairs in the bulk of the material [12-16]. Indeed, it is very difficult to create and stabilize a single type of defect in a compound semiconductor because the very creation of the defect changes the Fermi level of the system and thereby alters the concentration of the other types of defects. Multiple defects, after being generated, may form cluster and/or pair to lower the local free energy and get stability. Most recent reports show enhanced thrust on this agenda to unearth real nature and energetics associated with the formation of defects in ZnO (ref. 12). The issue becomes more challenging when impurities are incorporated in the system, intentionally [17, 18] or unintentionally [19]. Before presenting our results a survey of interesting outcomes on this agenda should be discussed here. Very recently, neutral $V_{Zn}$–$V_O$ (zinc-oxygen divacancies) have been detected in electron irradiated ZnO single crystal [20]. Such pairs dissociate when the crystal is annealed above 250 °C [20]. In fact, it has been shown that during growth, $V_{Zn}$ (zinc vacancy) and $V_O$ (oxygen vacancy) simultaneously appear in the system[21] and most probably they would remain as nearest neighbour divacancies. The nature of open volumes due to $V_{Zn}$ or $V_{Zn}$–$V_O$ defects has been estimated theoretically by positron annihilation lifetime (PAL) calculation [22] which closely matches with the experimental findings [8]. Chen et al. also identified[17] clustering of several (4-6) $V_{Zn}$–$V_O$ divacancies from the measured value of PAL in Al irradiated ZnO. Agglomeration of $V_{Zn}$ and $V_O$ defects (in general $(V_{Zn})_m - (V_O)_n$ type with m ≈ n) during annealing (at relatively lower temperature [23]) or by multiple collision cascades of energetic ions may lead to nano-voids in the system [24, 25]. From surface energy consideration, if two open volume defects with similar radii coalesce to a form larger one, energy is lowered roughly by 20%. Such larger open volumes are more stable and thermally recover in different stages above 300 °C (refs. 8,12). Most probable agglomerated form of vacancies in ZnO at different annealing temperatures has been investigated using kinetic Monte Carlo simulation[12]. The study predicts strong ability of $V_{Zn}-V_O$ clusters to attract monovacancies such as $V_{Zn}$s and $V_O$s. On the contrary, strong attractive interaction between two donor defects such as $V_O$ and $I_{Zn}$ (interstitial zinc) in ZnO have been predicted by a density functional theoretical calculation using hybrid functionals[13]. In an ion irradiated system all possible vacancies and interstitial defects are incorporated. A possibility may occur that both type of stable defect pairs $I_{Zn}-V_O$ and $V_{Zn}-V_O$ reside closely and can give rise to ~ 1.9 eV photoluminescence (PL) peak (red PL) at cryogenic temperatures [26]. Similar red PL emission in the photon energy range 1.6-2.1 eV have been found in ion implanted ZnO single crystal [15]. The authors attributed the origin of such PL peaks to the large (at least 3-4 $V_{Zn}$) and small size $V_{Zn}$ clusters. However, presence of $V_O$ in such vacancy clusters has not been ruled out. In fact, recent theoretical works envisage that broad red luminescence can be due to $V_{Zn}-V_O$ (ref. 16 & 27). In presence of H, Li, Al, N and other external impurities, several new type of vacancy-impurity complexes such as $Li_{Zn}-V_{Zn}$, $I_{Li}-N_O-V_{Zn}$, $N_O-V_{Zn}$, $V_{Zn}-N_O-H^+$, $Al_{Zn}-V_{Zn}$ etc. become most probable defects [28-32]. New and fascinating opto-electronic properties of ZnO have been theoretically predicted[3,16,31] and experimentally explored[14,19,29,32] using native vacancy clusters as well as their effective interaction with impurity atoms.

The present report aims to provide correct and comprehensive understanding on the evolution of defective state in granular ZnO with increasing fluence of 1.2 MeV Ar ion beam.

X-ray photoelectron spectroscopy (XPS), PL (at low temperature) and Raman spectroscopy (at room temperature) have been employed to investigate the nature of defects/defect clusters after the irradiation. It has been earlier observed that huge resistance reduction takes place along with the red-brown colouration due to 1.2 MeV Ar ion irradiation[33]. It has also been observed that resistance reduction takes place due to high temperature annealing, however, only yellowish colouration appears in ZnO but no red-brown colouration [8]. It can be understood that additional absorption in the green region (along with blue region, ~ 410 nm) of the spectrum is responsible for the change in colour from yellow to red-brown. The defect species responsible for absorption near 410 nm is still a matter of investigation. Energy level due to neutral $V_O$s have been proposed as a source of photon absorption in this region [34]. Recent investigation shows that concentration of $V_O$s (in ZnO) is an important parameter for the photon absorption peak energy [3]. However, the increase of the concentration of $V_O$ will necessarily invite $I_{Zn}$ or $V_{Zn}$ defects from the requirement of local free energy minimization. It is clear that presence of $V_O$s is necessary but simply $V_O$ or their clusters may not be the source of colour in ZnO. The analysis using SRIM software [35] estimates that in ZnO, 1.2 MeV Ar ions induce roughly 60 times $V_O$s compared to that of the $V_{Zn}$s near the subsurface region[36]. This information is a guide for understanding the results of the present investigation with a focus on $V_O$ rich defect clusters. It is also to be noted here that energetic Ar ions deposit energy in the target via electronic excitations (electronic energy loss, $S_e$) and direct elastic collisions (nuclear energy loss, $S_n$) [33]. The knockout of target atoms from their lattice positions takes place due to $S_n$ only. $S_e$ induces excitation and ionization of target electrons and thereby causing defect re-organization to some extent. Higher $S_n$ produces more vacancy interstitial combination and higher $S_e$ promotes more interaction between the generated defects so that, after equilibrium, a new defective state of the material is evolved depending on the particular combination of $S_e$ and $S_n$.

## 2. Experimental outline

Commercially available ZnO powder (99.99%, Sigma-Aldrich, Germany) have been pelletized and then annealed in air for four hours at 500 °C. Annealed ZnO pellets have been irradiated with 1.2 MeV Ar ion beam with fluence 0 (ZnO-U), $1\times10^{14}$ (ZnO-IL) and $1\times10^{16}$ (ZnO-IH) ions/cm$^2$ using low energy ion beam facility (LEIBF) [37], IUAC, New Delhi. The name of the samples henceforth will be used, have been given in respective parenthesis.

Electrical characterizations are performed using Keithley 4200-SCS parameter analyzer. X-ray photoelectron spectroscopy (XPS) has been carried out using PHI 5000 Versa Probe II (ULVAC, PHI, Inc.) spectrometry, equipped with a microfocused (100 mm, 25 W, 15 kV) monochromatic Al K$_\alpha$ X-ray source of energy 1486.6 eV. The binding energy in the XPS data has been calibrated considering C1s peak position at 284.6 eV. The Raman measurements were performed using a micro Raman set-up consisting of a spectrometer (Lab RAM HR Jovin Yvon) and a Peltier cold CCD detector. A He-Ne laser with wavelength of 633 nm was used as an excitation light source. PL spectra in the range (10–300 K) have been taken using He-Cd laser source as an excitation source, operating at 325 nm with an output power 45 mW and a TRIAX 320 monochromator fitted with a cooled Hamamatsu R928 photomultiplier detector. Glancing angle X-ray diffractogram have been recorded using a Bruker D8 advance diffractometer with Cu K$_\alpha$ (1.54 Å) excitation.

## 3. Results and discussion

Figure 1 shows the core level XPS spectra of O1s (Fig. 1a-1c) and Zn2p (Fig. 1d) levels for all three samples. The O1s peaks for all samples bear a shoulder at higher binding energy side as clearly seen in figure 1(a-c) which increases for irradiated sample with the increase of fluence. The O1s signals shown in figure 1(a-c) are deconvoluted into two Lorentzian peaks. Generally, the peak with lower binding energy (529.8 eV for ZnO-U sample) corresponds to $O^{2-}$ on wurtzite structure of ZnO and higher binding energy peak (531.2 eV) is attributed to $O^{2-}$ in the oxygen deficient regions within the host matrix. The weight of the component reflects the relative abundance of $V_O$s inside ZnO (ref. 38). It is clear from Fig. 1 that Ar irradiation produces stable $V_O$s in the subsurface region. Looking to the Zn related peaks, an overall shift is visible towards higher energy for ZnO-IL sample (Fig. 1d). The energetic difference between the peaks for Zn $2p_{3/2}$ (at 1020.8 eV) and for Zn $2p_{1/2}$ (at 1043.9 eV) is 23.1 eV (for ZnO-U) which indicates that zinc atoms are in + 2 oxidation (Zn-O bonding) state. The same peaks are at 1021.6 eV and 1044.5 eV, respectively for ZnO-IL sample. The energetic shift of these Zn$2p_{3/2}$ and Zn$2p_{1/2}$ (from 1020.8 to 1021.6 eV and from 1043.9 to 1044.5 eV) peaks indicates that Zn is in its more oxidized state after irradiation. So, more charge has been transferred from the Zn atom making the nucleus less shielded by the electron cloud which thereby increases a slight increase of Zn 2p electron binding energy. This is a typical case where Zn atom has to be displaced from its equilibrium lattice position with an oxygen interstitial ($I_O$) in the vicinity making an $I_{Zn}$-$I_O$ combination. Theoretically, Raman spectroscopic signature of such defect pair ($I_{Zn}$-$I_O$) should be found ~ 509 cm$^{-1}$ (ref. 39) which is also seen here (Fig. 2b)[40]. In a similar situation, Kayaci et al. proposed[41] that generation of $V_{Zn}$ is expected which with $V_O$ form larger cluster. So, presence of $I_{Zn}$s as well as some $V_{Zn}$s is very much likely[41]. In fact, + 0.4 eV shift for Zn $2p_{3/2}$ binding energy has been attributed for enhanced $V_{Zn}$ defects[42] in granular ZnO. Yun et al. have assigned the higher energy shift of Zn$2p_{3/2}$ spectra as a signature of $V_{Zn}$s in electron irradiated ZnO (ref. 43). The scenario seems to be little complex as the presence of $V_{Zn}$ clubbed with another defect (most probably $V_O$) is facilitating formation of $I_{Zn}$ species in the system. We argue that presence of $V_{Zn}$-$V_O$ (or $V_{Zn}$-$2V_O$) is important to induce local lattice deformation[31] and thereby shifting the nearest neighbour Zn atom from its equilibrium lattice position. This contention has been verified by another theoretical work in a different way[12]. Experimentally also, Rutherford backscattering (RBS/channelling) study reveals[44] that slightly displaced Zn atoms (0.07–0.1 Å) in ion irradiated ZnO can be the source of $I_{Zn}$ and thereby reducing resistance ~ $10^7$ orders in magnitude. No additional peak (Zn2p) shift has been observed with the increase of fluence. This indicates that the typical defect pair ($I_{Zn}$-$I_O$) is not energetically favourable to grow beyond certain concentration even if more disorder is incorporated. More and more generation of $I_{Zn}$, $I_O$, $V_{Zn}$ and $V_O$ defects with increasing fluence is being channelized to create different type of defect configuration as we shall see later.

To comment on the metallic Zn segregation, Zn(LMM) Auger transition spectra for all three samples have also been recorded and shown in figure 3. Each Zn (LMM) signal is resolved into two peaks by fitting Lorentzian distribution. The ZnO-IH spectra clearly shows that the weight of the lower binding energy peak (peak A in Fig. 3) has been notably increased compared to the ZnO-U and ZnO-IL spectra. The peak at lower energy side corresponds to the metallic Zn whereas the other (peak B in Fig. 3) is the signature of Zn-O bond of ZnO (refs. 45-48). Altogether, core level XPS and Auger spectra reveal monotonic increase of $V_O$ and $I_{Zn}$ type defects with the increase of fluence, however, combination of stable defect species may not be same in ZnO-IL and ZnO-IH samples. It will be more visible if one calculates Auger parameter $\alpha$ which indicates the overall chemical state of the corresponding atom[46]. From the Zn2p$_{3/2}$ and LMM spectra (Figs. 1 & 3), values of $\alpha$ parameter are 2009.815 eV (ZnO-U), 2010.070 eV (ZnO-IL) and 2009.624 eV (ZnO-IH). It is clear that both ZnO-IL and ZnO-IH bear $I_{Zn}$ defects but the coordination of oxygen atoms around Zn are quite different.

In the light of XPS findings, we would like to focus on the LTPL results of the present samples. 10 K PL spectra of near band edge (NBE) for the all the three samples have been shown in figure 4. The donor bound exciton (DBX) emission is observed at 3.365 eV (for ZnO-U & ZnO-IH) and at 3.369 eV (for ZnO-IL). The DBX peak intensity is largely affected due to ion irradiation[26] (as is also seen here) and annealing at elevated temperatures[19, 23]. So, the related shallow donor is intrinsic in nature and most probably originates from interstitial Zn ($I_{Zn}$) defects[1]. The thermal quenching of the DBX peak is plotted in figure 5. The activation energies ($E_a$) for quenching are in the range 8.5 – 10.1 meV for all the samples. Such a value of $E_a$ nicely correlates with $I_{Zn}$ related neutral donor defects[49]. Further, the red shifting of the DBX peak with the increase of temperature has been fitted (Fig. 6) with the equation given by Manoogian and Woolley[50], $E_{DBX}(T) = E_{DBX}(0) - UT^s - 2V\theta/(\exp(\theta/T) - 1)$, assuming that donor energy level follows the thermal shrinkage of the band gap[5]. In the above equation, $E_{DBX}(0)$ represents the DBX energetic position at 0 K, $U$ and $V$ represent the lattice dilatation and electron–phonon interaction term respectively. $s$ is an exponent and $\theta$ is related with the Debye temperature[4]. In both the figures 5 & 6, effects of 700 keV O ion irradiation[26] has been provided for comparison. The closeness of the parameters (shown in Figs. 5 & 6) indicates that no new donor state has evolved due to Ar ion irradiation. The inset of figure 4 shows deep level emissions (DLE) at 10 K. The broad defect luminescence in the range 2.7-3.0 eV is a characteristic of semi-insulating ZnO (ref. 51) with high concentration of compensating acceptor defects[52]. Most probable nature of such acceptors are $V_{Zn}$ and $O_{Zn}$s[26,28]. Consequently, the sheet resistance of ZnO-IL sample has increased an order in magnitude compared to that of the ZnO-U (Fig. 7). So broad PL in the range 2.7-3.0 eV can be donor acceptor type transition from extended $I_{Zn}$ states (due to different vacancy decoration of $I_{Zn}$ sites) to the $V_{Zn}$ levels[41, 53], assuming neutral $V_{Zn}$ energy level to be just above the conduction band. The PL emission ~ 2.4 eV (figure 4, inset) can be considered as stable $V_O^+$ in the bulk[2,4,33]. More abundant $V_O$s in ZnO-IL sample is in agreement with the XPS findings. The increase of fluence from $10^{14}$ to $10^{16}$ lowers DLE (and also overall PL) indicates incorporation of more non-luminescent centers. It can be understood that the size of the vacancy clusters[54] increases as collision cascades spatially overlap with the increase of fluence. More and more open volume defects merge to lower local free energy and as a result effective active volume of luminescent material decreases. Interesting enough, the DBX peak intensity is enhanced compared to ZnO-U due to highest fluence of irradiation (Fig. 4, sample ZnO-IH). The possible reason has been discussed afterwards with a model on irradiation effects in radiation hard materials.

Results of Raman spectroscopic measurement of the samples have been already shown in figure 2. Two strong non-polar Raman modes, $E_2$ (low) [101 cm$^{-1}$] and $E_2$ (high) [437 cm$^{-1}$] which are characteristic of wurtzite structure of ZnO along with two weak Raman modes, generally ascribed as $E_2$ (high) - $E_2$ (low) [331 cm$^{-1}$] and $A_1$(TO) [380 cm$^{-1}$] have been observed in the spectrum of ZnO-U sample. In ZnO, $E_2$ (low) Raman mode is related to Zn sub-lattice while vibration of oxygen atoms is responsible for the $E_2$ (high) mode[55]. Broad and asymmetric tailing of $E_2$ (low) Raman mode at higher frequencies is common in ion irradiated[26] or mechanically milled[56] ZnO systems. Tailing of $E_2$ (low) mode is very much pronounced in ZnO-IH Raman spectrum. Tentative origin of such tailing has been assigned with enhanced Zn sub-lattice disorder due to combination effect of $V_{Zn}$ and $V_O$ (ref. 57). In our view, the extent of tailing of $E_2$ (low) is correlated with the size of open volume defects. Higher disorder induces a loss of long range crystalline symmetry and hence decrease of intensity of $E_2$ (high) Raman mode is clearly observed. Further, irradiation causes evolution of another broad Raman mode (520-600 cm$^{-1}$) which is commonly attributed to $V_O$ and/or $I_{Zn}$ type defects i.e., oxygen deficient disordered state[55, 58]. The deconvolution of the broad Raman mode gives three different peaks centered ~ 540, 560 and 575 cm$^{-1}$ (Fig. 8). The peak ~540 cm$^{-1}$ is basically a second order Raman mode [$2B_1^{low}$; 2LA] which is normally activated in presence of defects [56]. It should be mentioned that $A_1$(LO) and $E_1$(LO) modes are very closely spaced[55] around 580 cm$^{-1}$. Here, 575 cm$^{-1}$ peak has the contribution of both A$_1$(LO) and E$_1$(LO) modes among them E$_1$(LO) is very sensitive to lattice defects, particularly to V$_O$s[55]. On the other hand, ~ 560 cm$^{-1}$ peak is related with the I$_{Zn}$ defects (ref. 59). However, it is more convincing to assign this Raman mode with the disorder at the interface of embedded Zn in a ZnO matrix[53]. Relative enhancement of 560 cm$^{-1}$ Raman mode has been observed for ZnO-IH compared to ZnO-IL sample (Fig. 8 inset). In this context, it is plausible to assign 575 cm$^{-1}$ Raman mode with the bulk V$_O$s in the ZnO crystal. The peak shapes become blurred in ZnO-IH sample indicating homogenized disorder[26] to some extent. Altogether, Raman spectroscopic investigation indicates clustering of individual vacancies for highest fluence of irradiation. At the same time, increase of embedded Zn atomic sites in ZnO lattice can also be understood (for ZnO-IH). This is conceivable if two types of defective zones gradually get separated[60] as more disorder is incorporated in the system, one leads to formation of nano-voids (agglomerated V$_{Zn}$-V$_O$ s) and the other transforms to highly non-stoichiometric Zn rich ZnO.

Based on the results discussed above we have proposed (schematically shown in fig. 9) a model which can be understood as a general feature of accumulation of disorder in a radiation hard material. In the low fluence regime, only scattered point defects appear in the system (Fig. 9a). Both V$_{Zn}$ and V$_O$s stabilize in the bulk of the material. V$_{Zn}$s are compensating acceptors in ZnO and hence "kill" donor electrons. V$_O$s are deep donors in ZnO and are unable to generate conductivity at room temperature[1, 16]. So incorporation of isolated V$_{Zn}$ and V$_O$ should increase the ZnO material resistance. With the increase of fluence, collision cascades starts to overlap and average separation between point defects decrease. $S_e$ plays an key role in the vacancy agglomeration process,[25, 61] most probably by promoting diffusion of vacancies. Energetically, the formation of V$_{Zn}$–V$_O$ divacancy (compared to distant V$_{Zn}$ and V$_O$)[31] or their higher order cluster[12] (such as (V$_{Zn}$)$_m$– (V$_O$)$_n$ type with m ≈ n, as mentioned earlier) are favourable in ZnO. In granular material, a fraction of such defects may migrate to the universal defect sink i e., grain boundaries (Fig. 9b). Increase of grain boundary depletion region width may additionally contribute to the enhancement of ZnO resistance. Further increase of fluence induces a characteristic change in the defective state of the material. Larger clusters agglomerate to generate voids which become new defect sink in the system. Deposited energy by ion beam in the form of $S_e$ and $S_n$ disseminate and/or recover a major fraction of the defects in the GB region. Residual defects migrate towards the voids increasing their size further. The system is now composed of large open volume defects (agglomerated $V_{Zn}$ and $V_{OS}$) and channels[25, 60] (Fig 9c) in between them which are Zn rich. Interstitial loops and vacancy zones nearby exist as seen by an in-situ high resolution transmission electron microscopic study of metallic Mg during ion irradiation[62]. If $I_{Zn}$ defects present in this nano-channels with sufficient number, conductivity should increase and an insulator to metal transition can also be induced[63]. In fact, the presence of embedded metallic zinc has been found[48] in laser irradiated black ZnO. Of course, the ratio of Zn/O in the subsurface region will depend on the relative sputtering of the elements during ion beam irradiation. In the channel region, interstitial dislocation loops as well as $V_{Zn}$, $V_{OS}$ should be present. However, overall disorder can be presumed to be less than compared to situation when GB regions exist in the sample. That is why XRD peak (e.g., 002) intensity is higher even for higher fluence (Fig. 10). Concerning the PL, voids reduce the optically active volume of the material, but the channels contribute to $I_{Zn}$ related DBX emission. Regarding colour of the sample, it has been conjectured that red-brown appearance is related with the Zn excess in ZnO (ref. 2). So $I_{Zn}$ defect dominated channel regions are the source of such colouration[33] in the high fluence irradiated ZnO. It has been shown earlier[33] that photon absorption ~ 400, 430 and 500 nm regions are responsible for red-brown colour of ZnO. Absorption of photons ~ 400 nm can only generate yellow coloured ZnO[34]. Additional absorption ~ 430 nm (2.88 eV) and 500 nm (2.48 eV) can arise due to presence of extended $I_{Zn}$ states[53] and $V_{Zn}$–$V_O$ defects[16,27] respectively. We believe, presence of $V_O$, $V_{Zn}$–$V_O$, $V_{Zn}$–$2V_O$ etc. in the disordered channel region contribute to the different absorption regions mentioned above. Consequently, lattice distortion due to such defects can stabilize[31,44] $I_{Zn}$ related shallow as well as semi-shallow donor states in this region.

Altogether, few interesting consequences can be foreseen based on the scenario presented here.

i) Amorphization is hardly achievable if a material can efficiently generate voids in application of energetic ion irradiation. Generation of voids depends on the diffusion coefficient of the concerned open volume defects (material property) and also on the sample temperature[60]. On the other hand, presence of foreign atom (chemical impurity, like As in GaSb alloy) can hinder void formation during ion irradiation and amorphization can be reached[61]. Till date, limited report exists in ZnO such as Si implantation induced secondary phase formation and amorphization [64].

ii) After reaching this new defective state, the overall defect as probed by RBS[65] or positron annihilation spectroscopy[15] will show a saturation of defects with increasing irradiation fluence further. This is because voids now act as both sink[11] and source of vacancies and the two distinctly different regions, as depicted in fig. 9c, are in defect equilibrium. In fact, a recent work envisages that buried extended defects can make ZnO more resistant for incorporating more disoder[66].

iii) Due to dominant $I_{Zn}$ species (donor defects) in the nano channels, it is impossible to achieve p-type conductivity in high fluence ($\geq 10^{16}$ ions/cm$^2$) irradiated ZnO. High concentration of implanted dopants like N (or Li) may form $N_2$ in voids (or form metal clusters) without being substituted at $V_{Zn}$ sites and generating $N_{Zn}$ ($Li_{Zn}$) acceptors.

iv) Ferromagnetism is expected in such highly disordered nano-network[7]. Indeed, room temperature ferromagnetic state has been detected in ZnO after high fluence of O (ref. 14) irradiation. This phenomenon is not limited to ZnO but has been found in highly disordered $SnO_2$ and $TiO_2$ (ref. 67) and other wide band gap semiconductors. However, here, the ferromagnetic state will not be Dietl like[68] (sufficiently high hole doped wide band gap ferromagnetic semiconductor), rather due to donor impurity band as proposed by Coey et al.[69] and observed by Zhang et al.[70], Maekawa et al.[14] and several other groups.

v) The change in the dominant ~ 3.315 eV transition in Fig. 4 (commonly termed as free to bound, FB transition[9, 26]) has not been discussed here. FB transition generally originates from localized acceptor defects in the vicinity of the GB[23]. If the new defective state (after high fluence of irradiation) evolves at the expense of grain boundary disorder, it is natural that FB peak intensity will be diminished compared to that of the DBX ($I_{Zn}$ related) as is noticed in Fig. 4.

vi) The modification of the disordered network due to annealing at relatively lower temperatures would be encouraging. Presumably, the system will not go back to its original granular nature and should show interesting physical properties, yet to be investigated in detail.

vii) The scenario in single crystalline ZnO will be different as such system has less in-built disorder (particularly GB regions) compared to that of granular materials. At this stage, it is not known how much fluence is required for the evolution (if at all) of two types of defective zones as sketched in fig 9c. Only it has been observed (not shown here) that for a fluence of $3\times10^{14}$ ions/cm$^2$ on ZnO single crystal (hydrothermally grown, MTI Corp., USA) sample lost its NBE drastically (similar to O irradiation effect[57]) even at 10 K. A recent report, however, shows a saturation[71] in defects above $10^{16}$ ions/cm$^2$ irradiation fluence of 300 keV Ar in ZnO single crystal target.

## 4. Conclusion

In this report, XPS, PL and Raman spectroscopic investigations have been combined in a general framework for understanding irradiation effects in granular ZnO. In low fluence regime when individual collision cascades do not spatially overlap, only scattered point defects are incorporated in the system. Adding disorder with higher fluence facilitates agglomeration of vacancies to larger clusters containing both $V_{Zn}$ and $V_{OS}$. Increase of vacancy cluster size further leads to appearance of voids in the system and the remaining regions becomes Zn rich. This is indeed a new defective state of the material as the voids have become major defect sinks not the grain boundary regions. Possible consequences of high fluence irradiation on the ZnO properties, particularly on colouration and conductivity, have been briefly outlined in the light of this model. Non-monotonic changes in the resistive nature of granular ZnO, often observed with increasing irradiation fluence, can also be addressed taking into account the evolution of changing defective nature of the system. To our knowledge, this work provides most general overview as of now for explaining and understanding ion induced disorder in ZnO.

**Conflicts of interest**

There are no conflicts of interest to declare.

**Acknowledgements**

Authors acknowledge the use of DST-FIST funded XPS facility at the Department of Physics, Indian Institute of Technology, Kharagpur. The authors are thankful to Dr. P. K.

Kulriya for extending support to glancing angle XRD facility at IUAC, New Delhi. The authors also thank Prof. S. K. Ray, Director, S. N. Bose National Centre for Basic Sciences, Kolkata, for helpful suggestions during PL data analysis. The authors acknowledge Dr. D. Sanyal, Variable Energy Cyclotron Center, Kolkata, for providing ZnO single crystal sample and Dr. Abhishek Nag, Indian Association for the Cultivation of Science, Kolkata, for illuminating discussion on the XPS results. S.P. acknowledges UGC, Government of India for providing the RFSMS fellowship (UGC/55/RFSMS/Physics) during his PhD tenure.

**References**

1. P. Camarda, F. Messina, L. Vaccaro, S. Agnello, G. Buscarino, R. Schneider, R. Popescu, D. Gerthsen, R. Lorenzi, F. M. Gelardia and M. Cannas, *Phys. Chem. Chem. Phys.*, 2016, **18**, 16237-16244.

2. J. Čížek, J. Valenta, P. Hruška, O. Melikhova, I. Procházka, M. Novotný and J. Bulíř, *Appl. Phys. Lett.*, 2015, **106**, 251902-4.

3. A. Catellani, A. Ruini and A. Calzolari, *J. Mater. Chem. C*., 2015, **3**, 8419-8424.

4. C. Drouilly, J-M Krafft, F. Averseng, S. Casale, D. B-Bachi, C. Chizallet, V. Lecocq, H. Vezin, H. L-Pernot and G. Costentin, *J. Phys. Chem. C*., 2012, **116**, 21297-21307.

5. A. Sarkar, M. Chakrabarti, D. Sanyal, D. Bhowmick, S. DeChoudhury, A. Chakrabarti, T. Rakshit and S. K. Ray, *J. Phys.: Condens. Matter*, 2012, **24**, 325503 (9 pp).

6. B. B. Straumal, A. A. Mazilkin, S. G. Protasova, A. A. Myatiev, P. B. Straumal, E. Goering and B. Baretzky, *Thin Sol. Films.*, 2011, **520**, 1192-1194.

7. R. Podila, W. Queen, A. Nath, J. T. Arantes, A. L. Schoenhalz, A. Fazzio, G. M. Dalpian, J. He, S. J. Hwu, M. J. Skove and A. M. Rao, *Nano Lett.*, 2010, **10**, 1383-1386.

8. S. Dutta, S. Chattopadhyay, A. Sarkar, M. Chakrabarti, D. Sanyal and D. Jana, *Prog. Mater. Sci.*, 2009, **54**, 89-136.

9. S. Guillemin, V. Consonni, L. Rapenne, E. Sarigiannidou, F. Donatini and G. Bremond, *RSC Adv.*, 2016, **6**, 44987-44992.

10. E. G. Barbagiovanni, R. Reitano, G. Franzò, V. Strano, A. Terrasi and S. Mirabella, *Nanoscale*, 2016, **8**, 995-1006.

11. A. Azarov, P. Rauwel, A. Hallén, E. Monakhov and B. G. Svensson, *Appl. Phys. Lett*, 2017, **110**, 022103-5.

12. J. Bang, Y-S. Kim, C. H. Park, S. Gao and S. B. Zhang, *Appl. Phys. Lett.*, 2014, **104**, 252101-5.

13. D. –H. Kim, G. –W. Lee and Y.-C. Kim, *Solid State Commun.,* 2010, **152**, 1711-1714.

14. M. Maekawa, H. Abe, A. Miyashita, S. Sakai, S. Yamamoto and A. Kawasuso, *Appl. Phys. Lett.*, 2017, **110**, 172402-5.

15. Y. Dong, F. Tuomisto, B. G. Svensson, A. Y. Kuznetsov and L. J. Brillson, *Phys. Rev. B*, 2010, **81**, 081201(R)-4.

16. A. Chakrabarty and C. H. Patterson, *J. Chem. Phys.*, 2012, **137**, 054709-5.

17. Z. Q. Chen, M. Maekawa, S. Yamamoto, A. Kawasuso, X. L. Yuan, T. Sekiguchi, R. Suzuki and T. Ohdaira, *Phys. Rev. B.*, 2004, **69**, 035210-10; Z. Q. Chen, M. Maekawa, A. Kawasuso, S. Sakai and H. Naramoto, *J. Appl. Phys.*, 2006, **99**, 093507-5.

18. A. Yu. Azarov, B. G. Svensson, A. Hallén, X. L. Du and A. Yu. Kuznetsov, *J. Appl. Phys.*, 2010, **108**, 033509-6.

19. S. Pal, T. Rakshit, S. S. Singha, K. Asokan, S. Duta, D. Jana and A. Sarkar, *J. Alloys Comp.*, 2017, **703**, 26-33.

20. M. S. Holston, E. M. Golden, B. E. Kananen, J. W. McClory, N. C. Giles and L. E. Halliburton, *J. Appl. Phys.*, 2016, **119**, 145701-7.

21. Y. J. Zeng, K. Schouteden, M. N. Amini, S. C. Ruan, Y. Lu, Z. Z. Ye, B. Partoens, D. Lamoen and C. V. Haesendonck, *ACS Appl. Mater. Interfaces*, 2015, **7**, 10617-10622.

22. G. Brauer, W. Anwand, W. Skorupa, J. Kuriplach, O. Melikhova, C. Moisson, H. von Wenckstern, H. Schmidt, M. Lorenz and M. Grundmann, *Phys. Rev. B.*, 2006, **74**, 045208-10.

23. H. Luitel, D. Sanyal, N. Gogurla and A. Sarkar, *J. Mater. Sci.*, 2017, **52**, 7615-7623.

24. K. S. Chan, L. Vines, L. Li, C. Jagadish, B. G. Svensson and J. Wong-Leung, *Appl. Phys. Lett.* 2015, **106**, 212102-5.

25. D. P. Datta, A. Kanjilal, S. K. Garg, P. K. Sahoo, D. Kanjilal and T. Som, *J. Appl. Phys.*, 2014, **115**, 123515-7.

26. S. Pal, A. Sarkar, P. Kumar, D. Kanjilal, T. Rakshit, S.K. Ray and D. Jana, *J. Lumin.*, 2016, **169**, 326-333.

27. Y. K. Frodason, Master's Thesis, University of Oslo, 2016.

28. J. Lv. and Y. Liu, *Phys. Chem. Chem. Phys.*, 2017, **19**, 5806-5812.

29. J. G. Reynolds and C. L. Reynolds, *Adv. Cond. Matt. Phys.*, 2014, **2014**, 457058-15.

30. J. E. Stehr, W. M. Chen, B. G. Svensson and I. A. Buyanova, *J. Appl. Phys.*, 2016, **119**, 105702-5.

31. A. Sarkar, D. Sanyal, S. Dechoudhury, D. Bhowmick, T. Rakshit and A. Chakrabarti, *Nucl. Instrum. & Meth. B*, 2016, **379**, 18-22.

32. B. Y. Zhang, B. Yao, Y. F. Li, A. M. Liu, Z. Z. Zhang, B. H. Li, G. Z. Xing, T. Wu, X. B. Qin, D. X. Zhao, C. X. Shan and D. Z. Shen, *Appl. Phys. Lett.,* 2011, **99**, 182503-3.

33. S. Chattopadhyay, S. Dutta, D. Jana, S. Chattopadhyay, A. Sarkar, P. Kumar, D. Kanjilal, D. K. Mishra and S. K. Ray, *J. Appl. Phys.*, 2010, **107**, 113516-8; S. Chattopadhyay, S. Dutta, P. Pandit, D. Jana, S. Chattopadhyay, A. Sarkar, P. Kumar, D. Kanjilal, D. K. Mishra and S. K. Ray, *Phys. Stat. Sol. C*, 2011, **8**, 512-515.

34. L. A. Kappers, O. R. Gilliam, S. M. Evans, L. E. Halliburton and N. C. Giles, *Nucl. Instrum. Meth. B*, 2008, **266**, 2953-2957.

35. J. F. Ziegler, J. P. Biersack and U. Littmerk, *Stopping Power and Ranges of Ion in Matter*, Pergamon, New York, 1985.

36. Here we have provided the estimate of vacancies for subsurface (~ 50 nm) region only. Mainly this region is being probed by XPS, PL and Raman spectroscopic techniques. Also, it has been assumed that ~ 99 % of the generated $V_{Zn}$s get recovered immediately after the generation. It should be remembered that SRIM software estimates only generated vacancies, not their corresponding numbers after stabilization (~ $10^{-8}$s after and more).

37. P. Kumar, G. Rodrigues, U.K. Rao, C.P. Safvan, D. Kanjilal and A. Roy, *Pramana*, 2002, **59**, 805-809.

38. F. Fan, P. Tang, Y. Wang, Y. Feng, A. Chen, R. Luo and D. Li, *Sensors and Actuators B: Chemical*, 2015, **215**, 231-240.

39. F. Friedrich, M. A. Gluba and N. H. Nickel, *Appl. Phys. Lett.,* 2009, **95**, 141903-3.

40. Not shown here, weak Raman mode ~ 510 cm$^{-1}$ has also been observed in ZnO single crystal after Ar ion irradiation (Fluence: $3\times10^{14}$ ions/cm$^2$).

41. F. Kayaci, S. Vempati, I. Donmez, N. Biyikliab and T. Uyar, *Nanoscale*, 2014, **6**, 10224-10234.

42. R. K. Sahu, K. Ganguly, T. Mishra, M. Mishra, R.S. Ningthoujam, S.K. Roy, L.C. Pathak, *J. Colloid Inter. Sci.*, 2012, **366**, 8-15; Q. Wang, Y. Yan, Y. Zeng, Y. Lu, L. Chen and Y. Jiang, *Sci. Rep.*, 2016, **6**, 27341-10.

43. E-J. Yun, J. W. Jung, Y. H. Han, M-W. Kim and B. C. Lee, *J. Appl. Phys.,* 2009, **105**, 123509-6.

44. T. Oga, Y. Izawa, K. Kuriyama, K. Kushida and A. Kinomura, *J. Appl. Phys.*, 2011, **109**, 123702-5.

45. Z. Yao, K. Tang, J. Ye, Z. Xu, S. Zhu and S. Gu, *Opt. Mater. Exp.*, 2016, **6**, 2847-2856; K. Tang, R. Gu, S. Zhu, Z. Xu, Y. Shen, J. Ye and S. Gu, *Opt. Mater. Exp.*, 2017, **7**, 1169-1179.

46. M. Krzywiecky, L. Grządziel, A. Sarfraz, D. Iqbal, A. Szwajca and A. Erbe, *Phys. Chem. Chem. Phys.*, 2015, **17**, 10004-10013.

47. D. Tainoff, B. Masenelli, O. Boisron, G. Guiraud and P. Mélinon, *J. Phys. Chem C*, 2008, **112**, 12623-12627.

48. E. H. Khan, S. C. Langford, J. T. Dickinson, L. A. Boatner and W. P. Hess, *Langmuir*, 2009, **25**, 1930-1933.

49. J. Sann, J. Stehr, A. Hofstaetter, D.M. Hofmann, A. Neumann, M. Lerch, U. Haboeck, A. Hoffmann and C. Thomsen, *Phys. Rev. B*, 2007, **76**, 195203-6.

50. A. Manoogian and J. C. Woolley, *Can. J. Phys.*, 1984, **62**, 285-287.

51. Z. Zhang, K. E. Knutsen, T. Merz, A. Yu Kuznetsov, B. G. Svensson and L. J. Brillson, *Appl. Phys. Lett.*, 2012, **100**, 042107-3.

52. S. Ghose, A. Sarkar, S. Chattopadhyay, M. Chakrabarti, D. Das, T. Rakshit, S. K. Ray and D. Jana, *J. Appl. Phys.*, 2013, **114**, 073516-6; S. Chattopadhyay, S. K. Neogi, P. Pandit, S. Dutta, T. Rakshit, D. Jana, S. Chattopadhyay, A. Sarkar and S. K. Ray, *J. Lumin.,* 2012, **132**, 6-11.

53. H. Zeng, X. Ningab and X. Liab, *Phys. Chem. Chem. Phys.*, 2015, **17**, 19637-19642.

54. A. Zubiaga, F. Tuomisto, V. A. Coleman, H. H. Tan, C. Jagadish, K. Koike, S. Sasa, M. Inoue and M. Yano, *Phys. Rev. B*, 2008, **78**, 035125-5.

55. V. Russo, M. Ghidelli, P. Gondoni, C. S. Casari and A. Li Bassi, *J. Appl. Phys.*, 2014, **115**, 073508-10.

56. S. Ghose, T. Rakshit, R. Ranganathan and D. Jana, *RSC Adv.*, 2015, **5**, 99766-99774.

57. S. Pal, A. Sarkar, S. Chattopadhyay, M. Chakrabarti, D. Sanyal, P. Kumar, D. Kanjilal, T. Rakshit, S.K. Ray and D. Jana, *Nucl. Instrum. Meth. B*, 2013, **311**, 20-26.

58. Z. Wang, S. Su, F. Chi-Chung Ling, W. Anwand and A. Wagner, *J. Appl. Phys.*, 2014, **116**, 033508-7.

59. H. J. Fan, R. Scholz, F. M. Kolb, M. Zacharias, U. Gösele, F. Heyroth, C. Eisenschmidt, T. Hempel and J. Christen, *Appl. Phys. A: Mater. Sci. & Proc.*, 2004 **79**, 1895-1900.

60. H. Yasuda, A. Tanaka, K. Matsumoto, N. Nitta and H. Mori, *Phys. Rev. Lett.*, 2008, **100**, 105506-4; L. Museur, A. Manousaki, D. Anglos and A. V. Kanaev, *J. Appl. Phys.*, 2011, **110**, 124310-8.

61. H. S. Alkhaldi, P. Kluth, F. Kremer, M. Lysevych, L. Li, M. C. Ridgway and J. S. Williams, *J. Phys. D: Appl. Phys.*, 2017, **50**, 125101 (8 pp).

62. W. Xu, Y. Zhang, G. Cheng, W. Jian, P. C. Millett, C. C. Koch, S. N. Mathaudhu and Y. Zhu, *Nature Commun.*, 2013, **4**, 2288-6.

63. S. Brochen, G. Feuillet, J-L. Santailler, R. Obrecht, M. Lafossas, P. Ferret, J-M. Chauveau and J. Pernot, *J. Appl. Phys.*, 2017, **121**, 095704-7; F. Singh, B. Chaudhary, V. Kumar, R. G. Singh, S. Kumar and A. Kapoor, *J. Appl. Phys.*, 2012, **112**, 073101-4.

64. M. Jiang, D. D. Wang, Z. Q. Chen, S. Kimura, Y. Yamashita, A. Mori and A. Uedono, *J. Appl. Phys.*, 2013, **113**, 043506-7; S. O. Kucheyev, J. S. Williams, C. Jagadish, J. Zou, C. Evans, A. J. Nelson and A. V. Hamza, *Phys. Rev. B*, 2003, **67**, 094115-11.

65. E. Wendler, O. Bilani, K. Gärtner, W. Wesch, M. Hayes, F. D. Auret, K. Lorenz and E. Alves, *Nucl. Instrum. Methods B*, 2009, **267**, 2708-2711.

66. A. Azarov, E. Wendler, K. Lorenz, E. Monakhov and B. G. Svensson, *Appl. Phys. Lett.*, 2017, **110**, 172103-5.

67. A. Sarkar, D. Sanyal, P. Nath, M. Chakrabarti, S. Pal, S. Chattopadhyay, D. Jana and K. Asokan, *RSC Adv.*, 2015, **5**, 1148-1152; D Sanyal, M. Chakrabarti, P Nath, A Sarkar, D Bhowmick and A Chakrabarti, *J. Phys. D: Appl. Phys.*, 2013, **47**, 025001 (5 pp).

68. T. Dietl, H. Ohno, F. Matsukura, J. Cibert and D. Ferrand, *Science*, 2000, **287**, 1019-1022.

69. J. M. D. Coey, K. Wongsaprom, J. Alaria and M. Venkatesan, *J. Phys. D: Appl. Phys.*, 2008, **41**, 134012 (6 pp).

70. X. Zhang, Y. H. Cheng, L. Y. Li, H. Liu, X. Zuo, G. H. Wen, L. Li, R. K. Zheng and S. P. Ringer, *Phys. Rev. B*, 2009, **80**, 174427-6.

71. A. Turos, P. Jóźwik, M. Wójcik, J. Gaca, R. Ratajczak, A. Stonert, *Acta Mater.*, 2017, **134**, 249-256.

**Figure captions:**

Figure 1. (a-c) O1s and (d) Zn2p peaks of XPS spectra of ZnO-U, ZnO-IL and ZnO-IH samples. Coloured circles represent corresponding colour of the samples.

Figure 2. (a) Raman spectra of ZnO-U, ZnO-IL and ZnO-IH samples. ZnO-IL and ZnO-IH spectra have been vertically shifted for better view. (b) Magnified view of 500-610 cm$^{-1}$ Raman spectra region showing a new signal at ~509 cm$^{-1}$ (identified with arrow) in ZnO-IL spectra and broad Raman modes at 520-600 cm$^{-1}$ in both ZnO-IL and ZnO-IH spectra.

Figure 3. Zn (LMM) Auger electron spectra of ZnO-U, ZnO-IL and ZnO-IH samples. Peak A energetic position for ZnO-U is at 495.125 eV and the same for peak B is at 497.625 eV.

Figure 4. Near band edge photoluminescence spectra of ZnO-U, ZnO-IL and ZnO-IH samples at 10 K. Corresponding deep level emissions (DLE) have been shown in the inset. PL emission ~ 3.315 eV has been assigned as FB (free to bound) transition.

Figure 5. Thermal quenching of PL intensities of main DBX emission (3.365 eV) for ZnO-U, ZnO-IL and ZnO-IH samples. Similar thermal quenching behaviour of oxygen ion irradiated (700 keV O; 3×10$^{14}$ ions/cm$^2$) ZnO has been shown for comparison. The solid lines are fit to the equation $I(T) = I(0)/[1 + A\exp(-E_a/k_B T)]$, where $I(T)$ is the PL intensity at temperature $T$, $A$ is a constant and $E_a$ is the activation energy of quenching. Respective $E_a$ values have been shown inside the bracket.

Figure 6. DBX emission peak shift with temperature for ZnO-U, ZnO-IL and ZnO-IH samples. Similar plot corresponding to oxygen ion irradiated (700 keV O; 3×10$^{14}$ ions/cm$^2$) ZnO has been shown for comparison. Such form of variations is fitted with the formula given by Manoogian and Woolley[50] for band gap shrinkage with increasing temperature.

Figure 7. Variation of sheet resistance with increasing fluence for 1.2 MeV Ar irradiated ZnO. Inset shows *I-V* characteristics of the ZnO-U, ZnO-IL and ZnO-IH samples.

Figure 8. Deconvolution of broad Raman modes (520-600 cm$^{-1}$) in ZnO-IL and ZnO-IH spectra. Best fitting using three Lorentzian have been shown. The inset shows the relative change of Raman mode ~ 560 cm$^{-1}$ with respect to irradiation fluence (both axes in logarithmic scale). Here unirradiated sample has been assigned with $10^{10}$ ions/cm$^2$ irradiation fluence. The solid line is a guide to eye.

Figure 9. Model representing the effect of ion beam irradiation in granular ZnO material. Shape, scale and colours are qualitative and for illustration only. (a) Case of low irradiation fluence where only scattered point defects (white points) appear. For 1.2 MeV Ar ion on ZnO, low fluence represents $\leq 10^{12}$ ions/cm$^2$ (below which chance to overlap of collision cascades is less likely) which is very difficult to realize. (b) Moderate fluence regime where collision cascades frequently overlap and initiate agglomeration of point defects to form larger defect clusters. (c) High irradiation fluence (certainly $\geq 10^{16}$ ions/cm$^2$ for 1.2 MeV Ar ion on ZnO) induced void formation and probable modification of GB regions is schematically addressed (details in the text).

Figure 10. Glancing angle X-ray diffractogram (scan range 2$\theta$: 30º-40º) of ZnO-U, ZnO-IL and ZnO-IH samples.

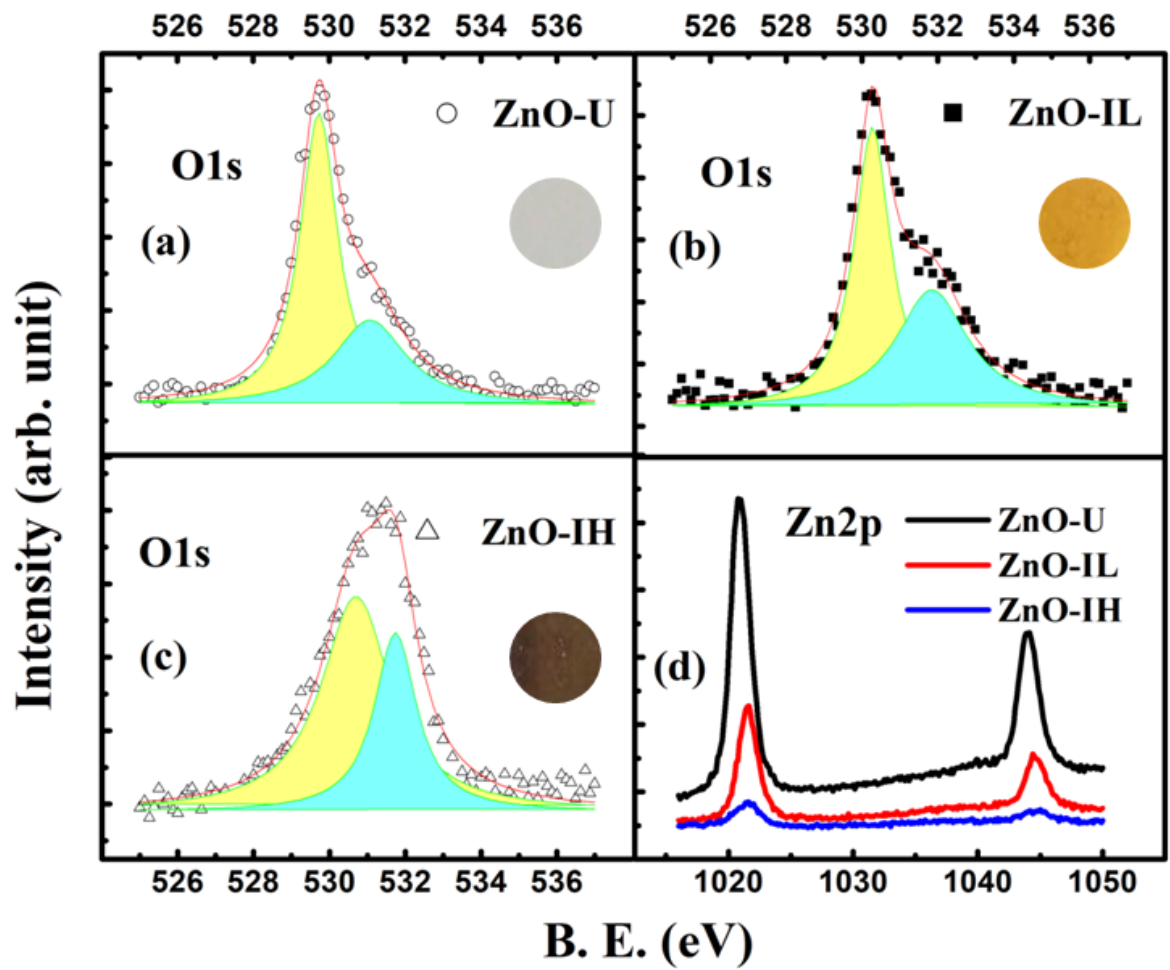

**Figure 1.**

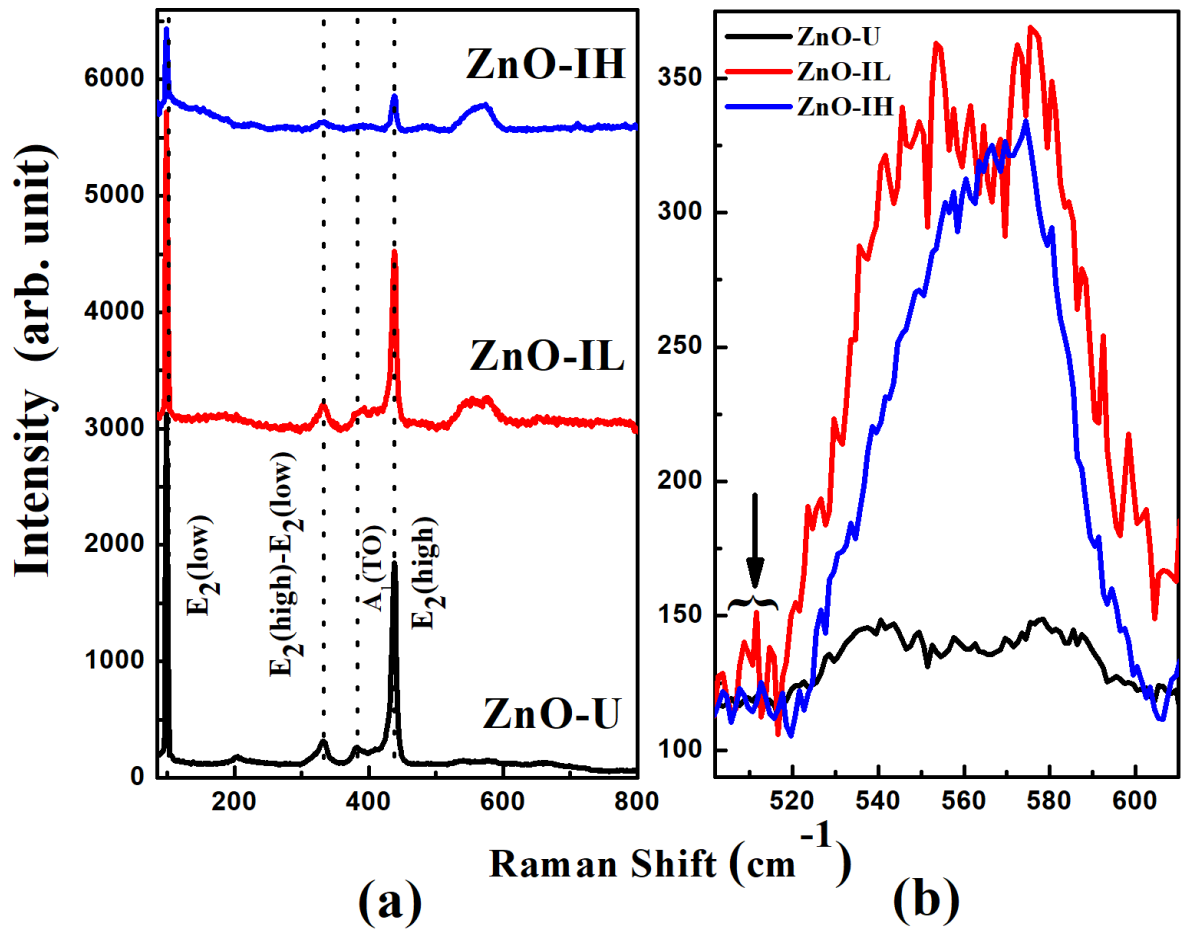

**Figure 2.**

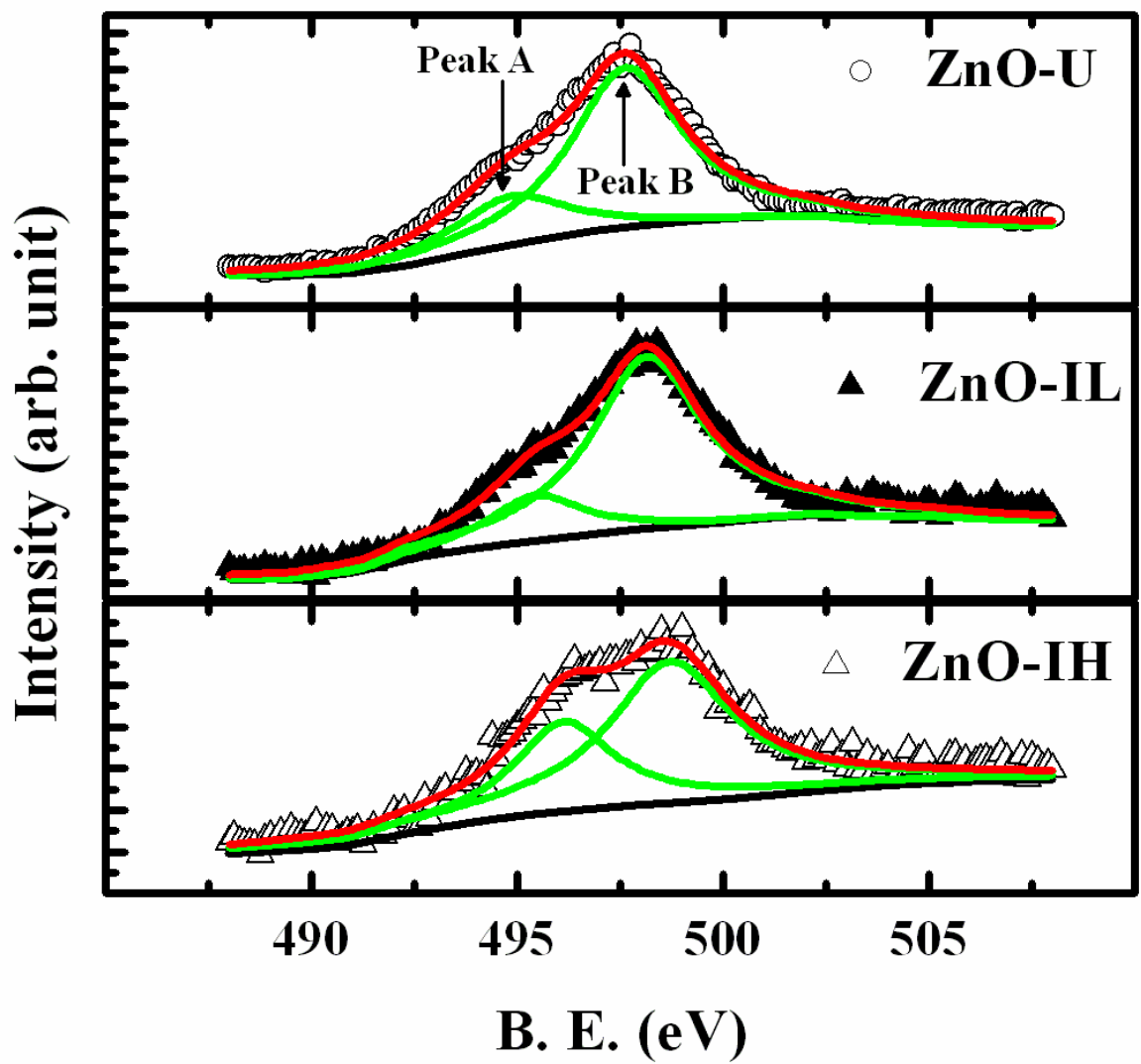

**Figure 3.**

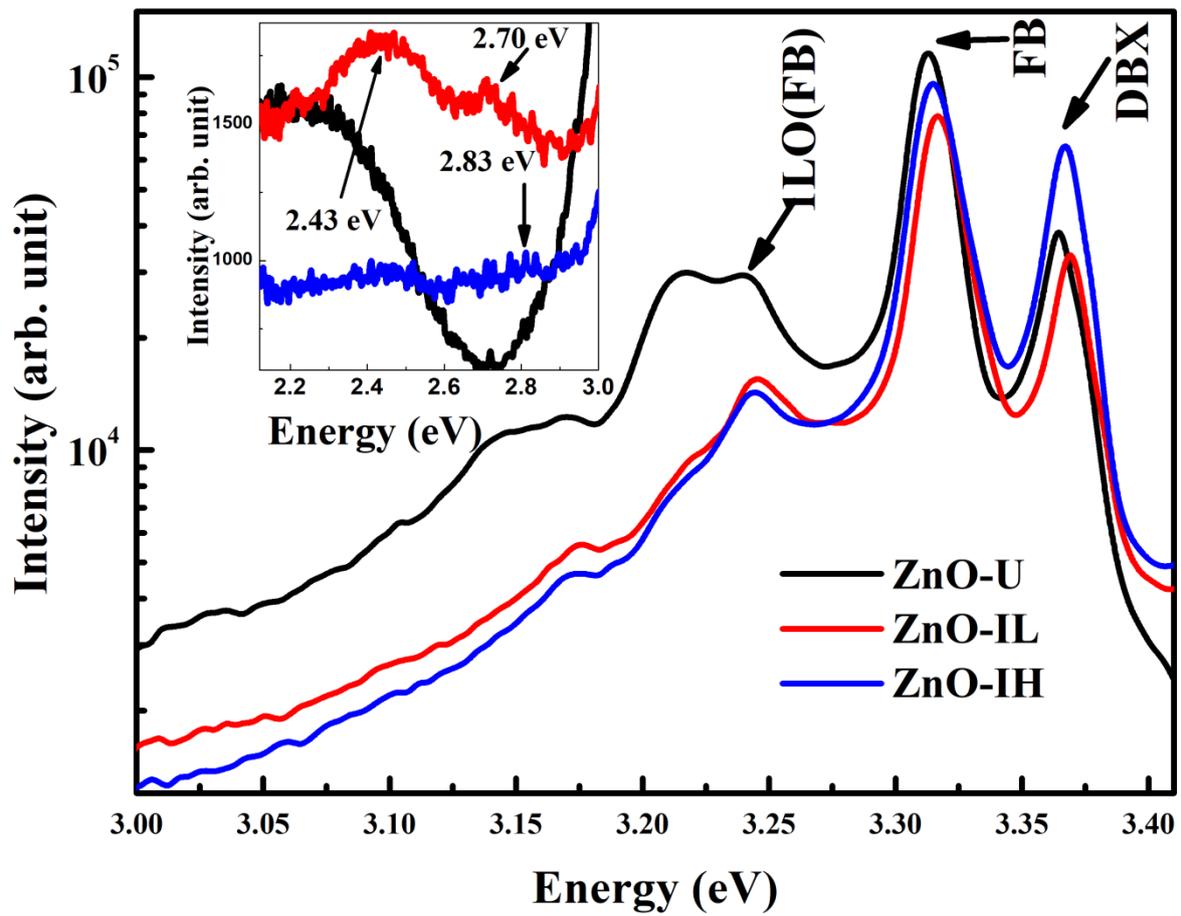

**Figure 4**

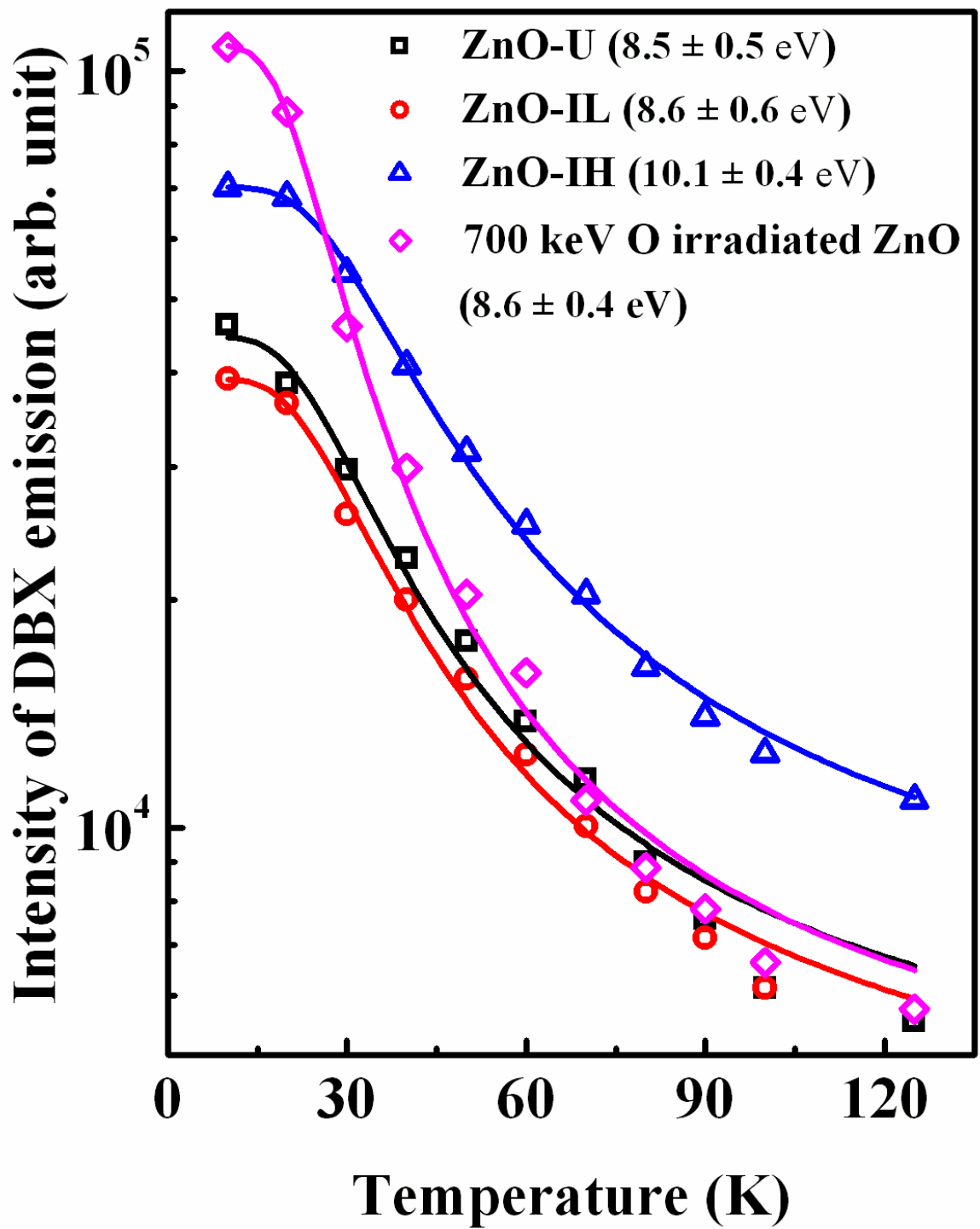

**Figure 5.**

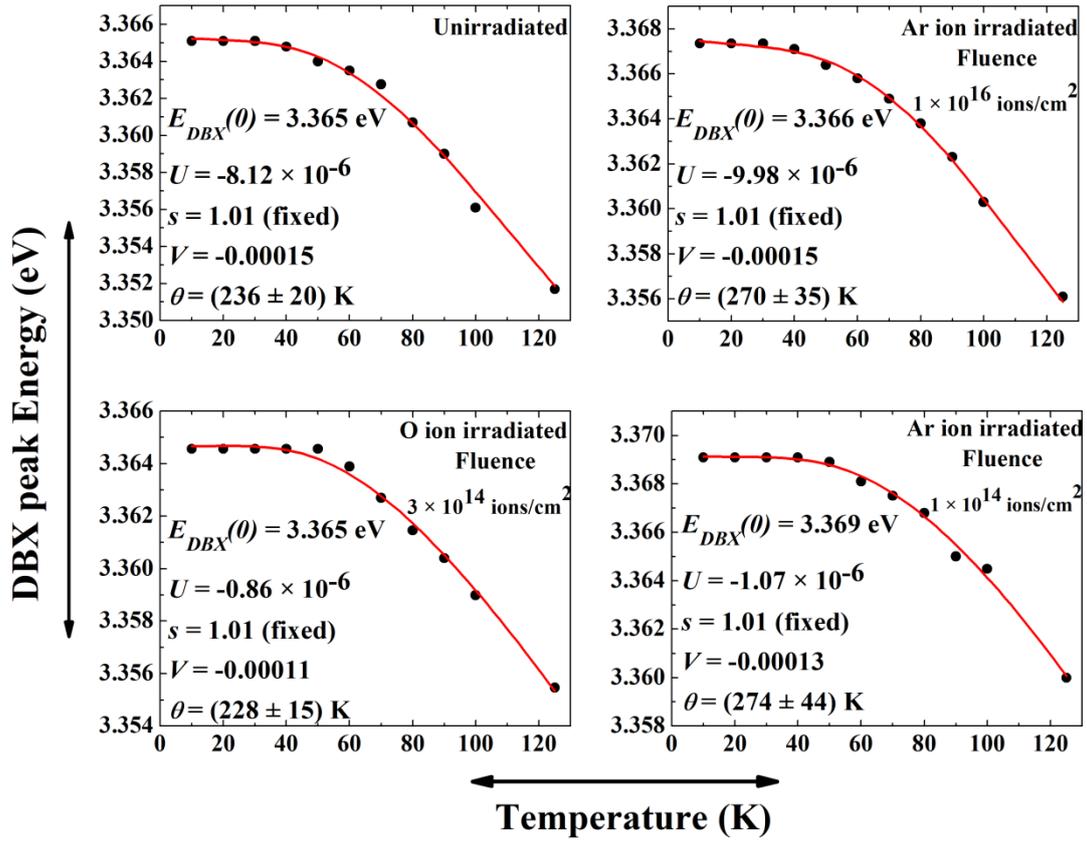

**Figure 6.**

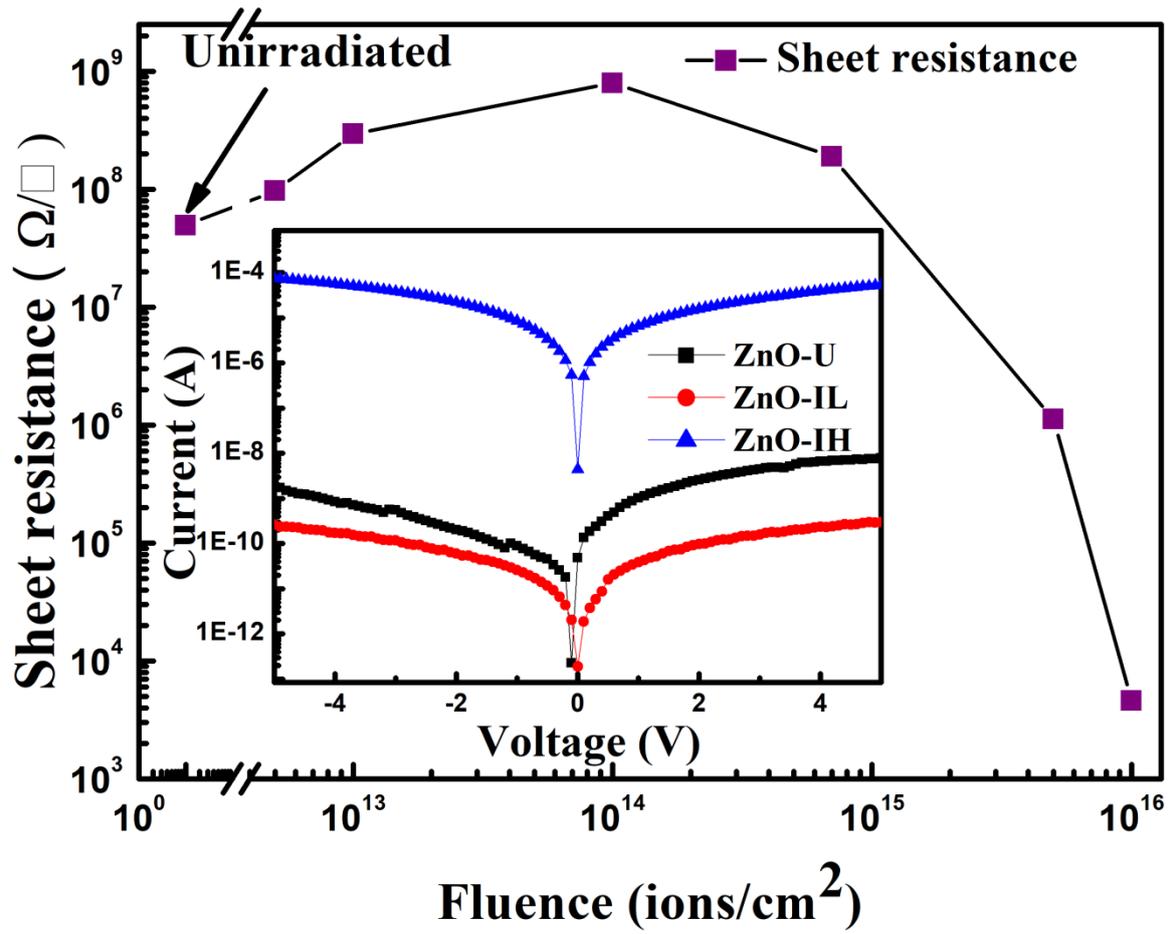

**Figure 7.**

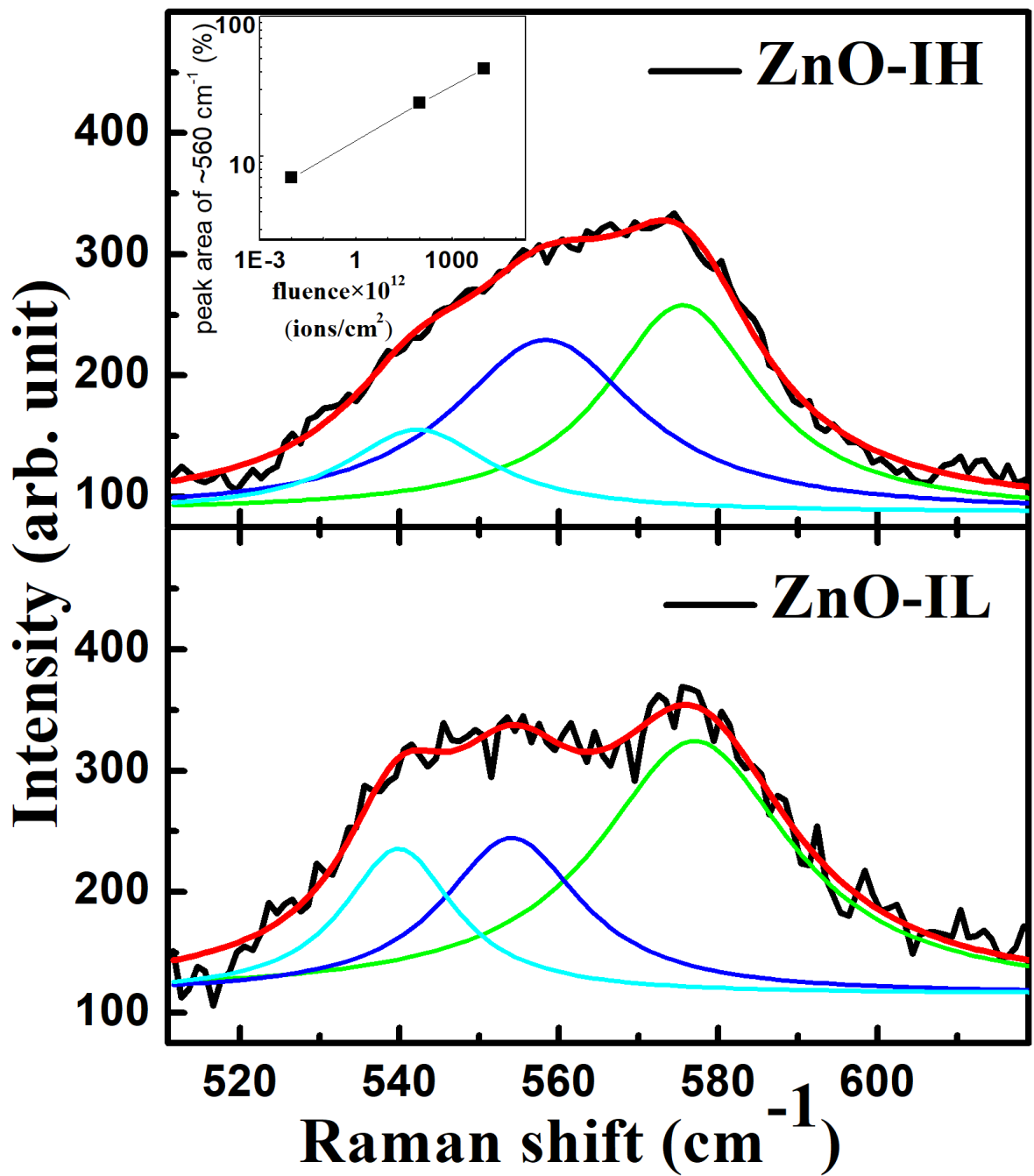

**Figure 8.**

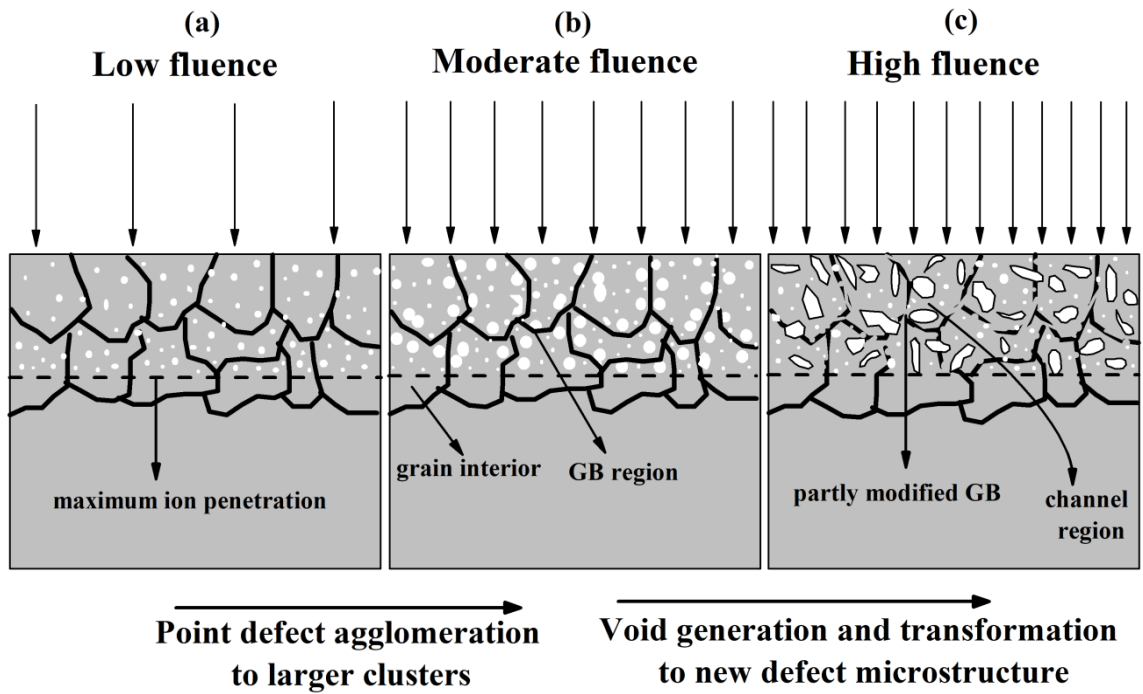

**Figure 9.**

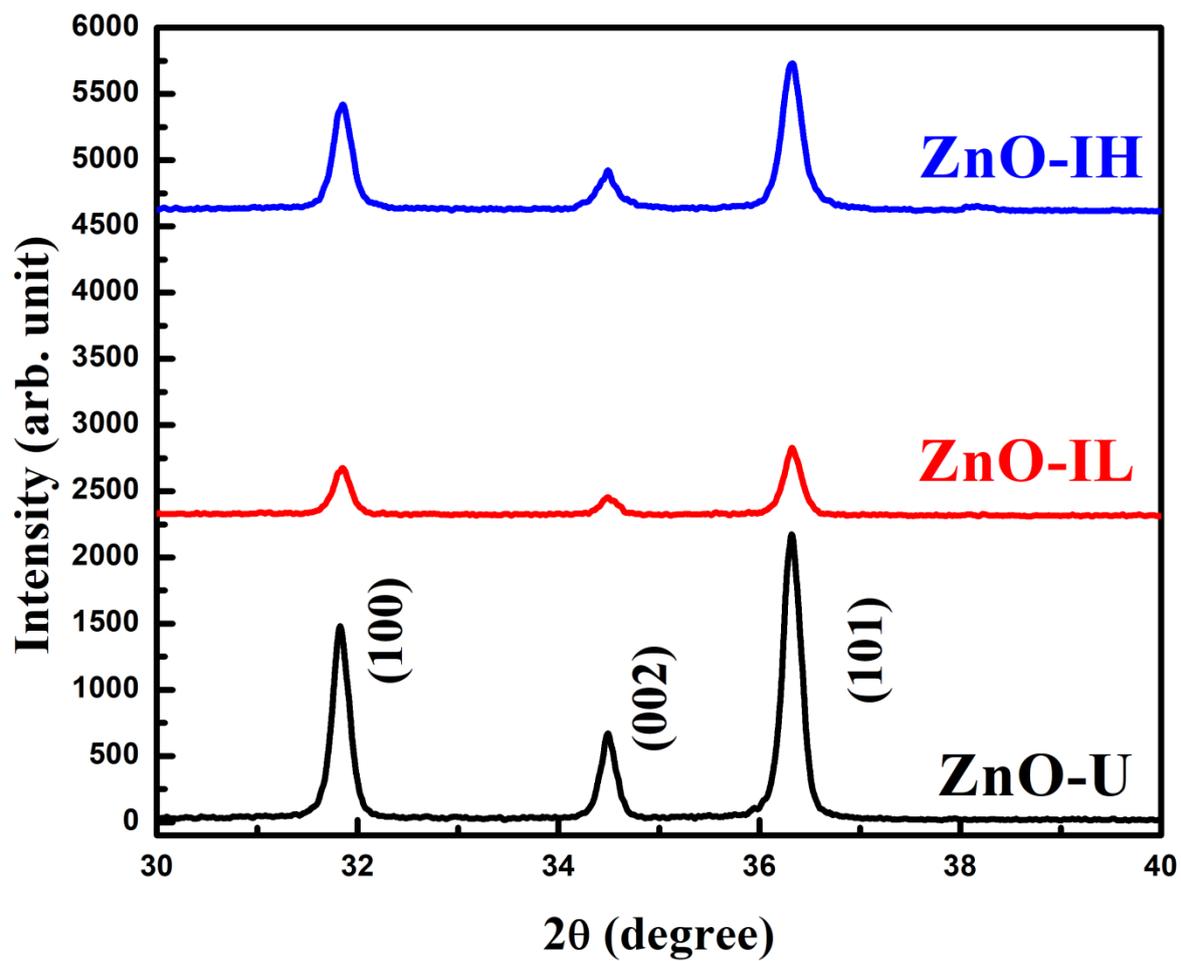

**Figure 10.**